\documentclass[preprint,preprintnumbers,amsmath,amssymb]{revtex4}
\usepackage{epsfig}
\usepackage{color}
\usepackage{newlfont}
\usepackage{verbatim}
\usepackage{amssymb,amsmath}

\begin{document}

\title{Proximity-Coupled Ti/TiN Multilayers for use in Kinetic Inductance Detectors}
\author{ Michael R. Vissers }
\email {michael.vissers@nist.gov}
  \author{Jiansong Gao}
  \author{ Martin Sandberg}
 \author{ Shannon M. Duff}
 \author{David S. Wisbey}
  \altaffiliation{Current address: Department of Physics, Saint Louis University, 3450 Lindell Blvd. Saint Louis, MO 63103, USA}
   \author{Kent D. Irwin}
 \author{David P. Pappas}
\email{david.pappas@nist.gov}
\affiliation{National Institute of Standards and Technology, \\ 325 Broadway, Boulder, CO, 80305}

\date{\today}

\begin{abstract}

We apply the superconducting proximity effect in TiN/Ti multi-layer films to tune the critical temperature, $T_C$,  to within 10 mK with  high uniformity (less than 15 mK spread) across a 75 mm wafer. Reproducible $T_C$'s are obtained from 0.8 - 2.5 K. These films had high resistivities, $\ge 100$ $\mu \Omega$-cm and internal quality factors for resonators in the GHz range on the order of 100k and higher. Both trilayers of TiN/Ti/TiN and thicker superlattice films were prepared, demonstrating a highly controlled process for films over a wide thickness range. Detectors were fabricated and showed single photon resolution at 1550 nm. The high uniformity and controllability coupled with the high quality factor, kinetic inductance, and inertness of TiN make these films ideal for use in frequency multiplexed kinetic inductance detectors and other potential applications such as nanowire detectors, transition edge sensors and associated quantum information applications.
 
\end{abstract}
\maketitle

Microwave kinetic inductance detectors (MKIDs) have seen rapid development over the past 10 years, with applications in astronomical instruments from sub-mm to gamma-ray\cite{Day-MKID} and non-astronomical applications\cite{Jonas-review, Cecil-arXiv}. The  key advantage of MKIDs is that they are based on high-Q resonators, and it is relatively simple to fabricate frequency-multiplexed devices from a single film of superconducting material. Titanium nitride (TiN), in particular, has recently attracted attention because resonators fabricated from this material have shown high kinetic inductance, normal resistivity, and very low loss for resonators in the GHz range\cite{JPL-TiN, Vissers-TiN}. This low loss translates into a very high internal quality factor, $Q_i$, while the high kinetic inductance and resistivity greatly improve responsivity and facilitate optical coupling, respectively.  Uniformity in these properties is therefore crucial for scaling up to large arrays of devices in astronomical applications.  

Another important property of TiN is the tunability of the superconducting critical temperature, $T_C$, from 0 to 5 K by adjusting the nitrogen concentration. This allows films made from various TiN$_X$ compounds to be tailored for specific applications. For example, films with  $T_C \sim 1~$K are the ideal choice for photon-counting MKIDs because the sensitivity improves as the $T_C$ is lowered (frequency shift per photon $\delta f \propto 1/T_C^2$), while it  is still high enough to prohibit thermal quasi-particle formation at the typical bath temperature of T$\sim100~$ mK. In addition, a $T_C$ of 1 K is needed for mm-wave MKIDs used in cosmic microwave background (CMB) detection. The two most used CMB bands for ground-based instruments are 90~GHz and 150~GHz, which require $T_C$ to be below $1.2~K$ and $2~K$, respectively, in order for the mm-wave photons to break Cooper-pairs in the superconductor \cite{Mazin_thesis}. In fact, TiN with $T_c\sim$1 K range is preferred for applications in the entire 90-300~GHz range. 

Sub-stoichiometric TiN films with $T_C$ around 1 K have been made into photon-counting detectors at UV/Optical/NIR wavelengths and have shown single photon sensitivity \cite{Gao_SubStoichometricTiN_arXiv, McHugh_SinglePhoton}. However, it is difficult to control the targeted $T_C$ in these sub-stoichiometric films; additionally, the films show large non-uniformity across the wafer\cite{Gao_SubStoichometricTiN_arXiv,Vissers-TiN-characterization-arxiv,JOLT}. This is due to two compounding effects. First, $T_C$ is a very strong function of the N content because the change in $T_C$ occurs over a narrow range of N concentration, just above to the Ti$_2$N phase. Second, it is difficult to maintain precise control of the nitrogen incorporation process from run-to-run and across large areas. This is due to to the fact that N is introduced to the process as a flow of gas into a reactive sputtering  chamber. In the chamber there is a complex interplay of gas flows, target voltages, and target erosion profiles.  By use of a typical UHV vacuum chamber with 75 mm sputter targets, variations on the order 500 mK between runs and across the wafer are typical\cite{Vissers-TiN-characterization-arxiv}. This difficulty in controlling $T_C$ is a great challenge that must be solved before TiN can be used in large MKID arrays. 

In this letter, we take a different approach. We use a multi-layer of pure Ti and stoichiometric TiN, exploiting the superconducting proximity effect to obtain the target $T_C$ while maintaining the desirable qualities of the TiN at the surfaces and interfaces\cite{Sandberg-etch, Wenner-TLS}. This approach works well because  it is comparatively easy to reproducibly grow layers of these materials that are homogeneous and have a constant thickness (within 5\%) across the wafer. Titanium and TiN  have $T_C$'s of 0.4 and 5 K, respectively, and by adjusting the relative thicknesses of these layers it is possible to tune the $T_C$ between these two limiting values.  

Theories to explain the superconducting proximity effect have been formulated since the first experiments were performed \cite{Meissner}.  The fundamental non-locality of superconductivity implies that for multi-layer thin films, $T_C$ depends not just upon the intrinsic superconducting properties of the constituent materials, but also on the interaction of the different layers.  Cooper's early description\cite{Cooper} of the proximity effect dealt with the $T_C$ of a normal-superconducting bilayer where the thicknesses of the two layers were less than any relevant length scale.  These same arguments can be extended to multi-layers \cite{Sato,Ventura,Luukanen}. The $T_C$ of these films is then determined by (1) the average of the $T_C$'s of the constituent parts weighted by the number of electrons contributed by each layer, and (2) the spatial extent of the proximitization. For the superconductor, this scale is determined by the size of the Gor'kov kernel, which in dirty films is $\propto (\xi l) ^{1/2}$\cite{Waldram}, where $\xi$ is the coherence length and $l$ the electron mean free path. Specifically, for TiN we have measured the Ginzburg-Landau coherence length $\xi_{TiN}=13\pm2$ nm (from measurements of $H_{C2}$ vs. temperature), and $l$, is on the order of the grain size, i.e. about 20 nm for these films.  For the normal metal, i.e. Ti well above its superconducting transition temperature, the proximitization scales with the electron mean free path, which also is limited by the grain size \cite{Reale_Ti-MFP}. This sets the Ti and TiN film thicknesses of interest for this study to be up to about 20 nm.

\begin{figure}
\includegraphics[width=10cm]{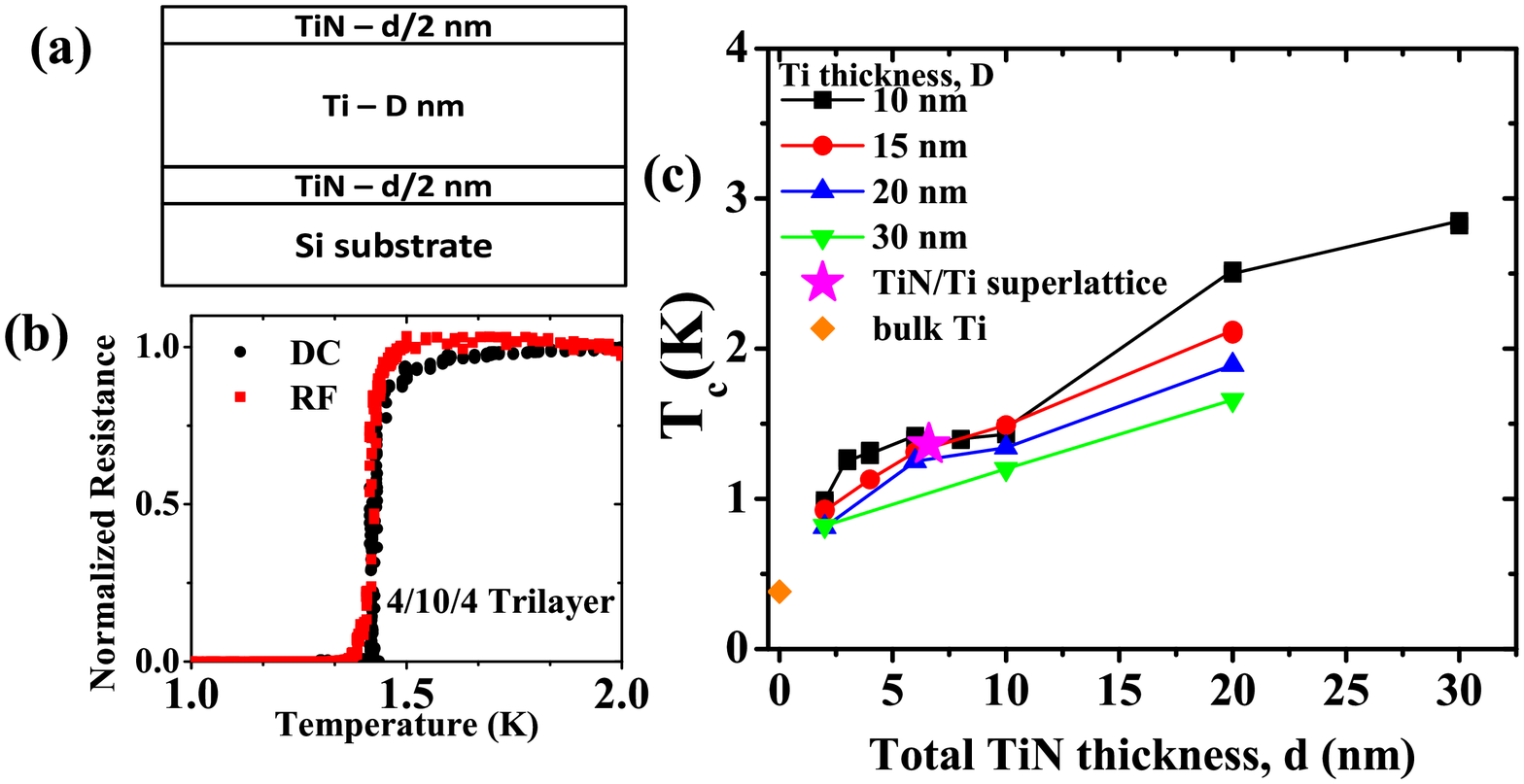}
\caption{(color on-line) (a) Schematic of a trilayer.  (b) Measured resistance versus temperature at DC and AC (6 GHz) of a 4/10/4 nm Ti/TiN/Ti trilayer.  (c) Measured $T_C$ at DC of trilayers vs TiN thickness for various Ti thicknesses.}
\label{Tc}
\end{figure}

We have chosen to create trilayer and superlattice structures with TiN at the top and bottom interfaces, e.g. TiN/Ti/$\cdot\cdot\cdot$/TiN. This preserves the good vacuum and Si-interface properties of TiN films. More specifically, it reduces frequency noise by limiting the number of two level systems (TLS's)\cite{JPL-TiN,Vissers-TiN} and suppresses oxidation by protecting the Ti film. The non-locality of superconductivity implies that the trilayers and multi-layers will act functionally the same as the bilayer structure discussed above, with a single $T_C$ for the entire film. The films were grown in a UHV sputter tool similar to that described previously\cite{Vissers-TiN}. Figure \ref{Tc}(a) illustrates the heterostructure of the deposited film. They were grown on high-resistivity Si wafers.  The wafers were first cleaned in HF, immediately transferred to the growth chamber, and then heated to 500 $^\circ$C on a rotating platen. A short (60 second) preliminary soak was performed in Ar:N$_2$ with the shutters on the gun and at the sample closed and the sample RF bias on. This soak serves to form a $\approx$1 nm layer of SiN that acts both as an insulator and as a buffer layer for subsequent growth of (200)-textured TiN. The buffer layer reduces the nucleation of the (111)-texture, which we have found deleterious for the RF properties of TiN\cite{Vissers-TiN}. The first layer of stoichiometric TiN was deposited at high temperature by use of reactive sputtering in the Ar:N$_2$ mixture. The N$_2$ partial pressure was stabilized using an upstream thermal mass flow controller and throttling the chamber pump downstream.  The sample was then cooled, and subsequent layers were grown at low ($\approx$300 K) temperature.  Between growths of the individual layers, the Ti sputter gun was run with both shutters closed and $N_2$ flow set either off or on to prepare the target surface for the next layer.  This is important because during the TiN deposition the surface of the Ti sputter source becomes fully nitrided and must be cleaned in order to deposit pure Ti, then after the Ti deposition the target needs to be re-nitrided.  

The trilayer and superlattice films were then patterned into test structures for measurements of $T_C$, resonator quality factor, and detectors for photon counting at near infra-red wavelengths.  For all measurements the samples were cooled in an adiabatic demagnetization refrigerator (ADR) that has a base temperature of 50 mK and variable temperature control up to 10 K.  Room temperature resistivities on the order of 100 $\mu\Omega-\textrm{cm}$ and a RRR of $\sim$1 were measured.  Critical temperature measurements were conducted at both DC and AC (at RF $\approx$ 4 GHz). As shown in Fig. \ref{Tc}(b) the DC transition from the normal to superconducting states is very sharp, with a width of less than 25 mK.

Figure \ref{Tc}(c) shows the measured superconducting transition at DC for films as a function of total TiN thickness, d, and Ti thicknesses, D.  As expected, increasing the TiN thickness (or decreasing the Ti thickness) leads to an increased $T_C$. Furthermore, the heavier weighting of the more metallic Ti is evident because the measured $T_C$'s are less than a simple average would predict.  

More significantly for this work, we also observe a range of TiN thicknesses where $T_C$ is relatively insensitive to the thicknesses of the constituent layers.  This region, with $T_C \approx$1.3 K and total TiN thickness between 5-10 nm exhibits a significantly reduced slope in $T_C$ vs TiN thickness.  The origin of this effect is still under investigation, but we hypothesize that it is due to inter-facial effects and intermixing between the TiN and the Ti that do not scale with TiN thickness. Regardless of the microscopic cause, this relative insensitivity of $T_C$ to film thicknesses enhances the reproducibility and homogeneity of the film properties. This carries over to the thicker superlattice films. Figure \ref{Tc}(c) also shows data taken from a 3/10/3/10/3/10/3/10/3 nm TiN/Ti/TiN superlattice (star symbol). With $T_C$=1.4 K, i.e. the same as the trilayers with similar Ti/TiN ratios, the superlattice illustrates the scalability of the proximity effect.  This result implies that films of almost arbitrary thickness can be grown, with the $T_C$ controlled by the relative thicknesses of the TiN and Ti layers. In addition, we find that $T_C$ for nominally identical films is reproducible within $\le 2\%$.  

\begin{figure}
\includegraphics[width=8cm]{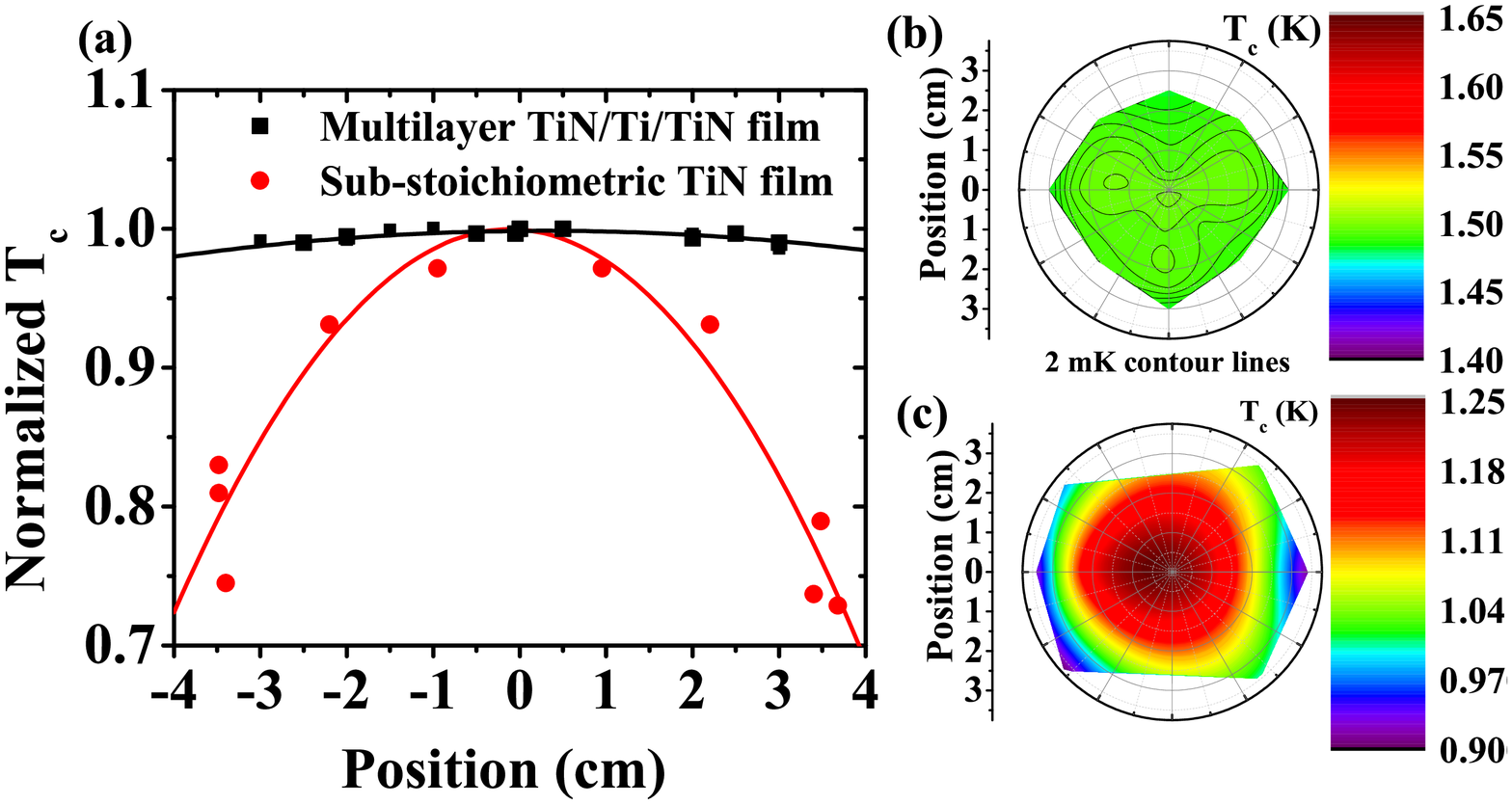}
\caption{(color on-line) (a) $T_C$ vs position for the multi-layer stoichiometric TiN/Ti/TiN(black) and mono-layer (red) sub-stoichiometric TiN based thin films.  The multi-layer has 10$\times$ less variation in $T_C$. The lines are a guide to the eye. (b) and (c) Corresponding contour plots of measured $T_C$ in stoichiometric multi-layer and sub-stoichiometric mono-layer wafers.}
\label{WaferMap}
\end{figure}

To illustrate and compare the homogeneity of $T_C$ for these multi-layer films on a given wafer relative to sub-stoichiometric films, we show the variation of $T_C$ across a 75 mm wafer for both processes in Fig. \ref{WaferMap}. As shown in Fig. \ref{WaferMap}(a), the multi-layer films show less than 1.5\% variation of $T_C$ across the wafer (15 mK variation on a $T_C$ = 1 K wafer). The wafer map shown in Fig. \ref{WaferMap} (b) illustrates that multi-layers are very homogeneous across the wafer, with only a slight radial dependence that is at the limit of the $T_C$ measurement resolution. The reproducibility of $T_C$ and better homogeneity across the wafer for the multi-layers is due to the fact that the absolute thickness of the Ti and TiN can be controlled precisely at the sub-nm level combined with the relative insensitivity of $T_C$ to TiN thickness in this range. 

For comparison, Fig. \ref{WaferMap}(a) shows that the $T_C$  on the wafer with sub-stoichiometric TiN varies by more than 25\%, while in Fig. \ref{WaferMap}(c) it can be seen that the non-uniformity is strongly radially dependent.  The non-uniformity is thought to be due to uneven nitrogen gas flow throughout the chamber. This will unevenly react with unbonded Ti at the substrate, and the radial dependence will be caused by the symmetry imposed by the rotation of the substrate during deposition. The extreme sensitivity of the TiN's $T_C$ to the nitrogen flow means that small variations in the nitrogen concentrations due to gas loading and inhomogeneous flow in the chamber lead to variations of  hundreds of mK in the TiN$_X$ $T_C$ across the wafer. \cite{Vissers-TiN-characterization-arxiv,JOLT,Gao_SubStoichometricTiN_arXiv}.   

For RF measurements, the films were patterned into quarter-wave co-planar waveguides (CPWs) and lumped-element kinetic inductance detector (LEKID)-style resonators. The resonators were capacitively coupled to a CPW  feed-line. We first measured the RF transmission through the feed-line near $T_C$. Figure \ref{Tc}(b) shows the transmission data from a device fabricated from a 4/10/4 nm TiN/Ti/TiN film with a $T_C$=1.4 K.  Since the film is much thinner than the penetration depths of the TiN ($\sim$300 nm) or Ti ($\sim$ 100nm)  \cite{Vissers-TiN},  the current should be distributed throughout the film thickness. Thus, the RF transmission is a good measure of the $T_C$ of the entire film stack. For example, if the Ti layer in the center of the multi-layer was too thick, hence not completely proximitized, the RF transmission would not stop decreasing until the full multi-layer was superconducting. We measured a sharp S$_{21}$ transition of the multi-layer at $T_C(DC)$ with no other structures at low temperature. This indicates that the all of the layers of the film are fully proximitized. 

We then performed standard resonator measurements at the ADR base temperature. The measured and derived properties for several resonators are listed in Table \ref{ComparisonTable}. These include $T_C$ at DC, $Q_i$, the penetration depth $\lambda$ (extracted from the kinetic inductance-induced frequency shift fraction or the 4 K resistivity\cite{Vissers-TiN}), and the normal state DC resistivity, $\rho_n$. The values for the multi-layers compare well with the sub-stoichiometric TiN film. A general trend of decreasing $Q_i$ is observed as $T_C$ decreases. However, these numbers should still be high enough for applications in a MKID.

In order to directly test the efficacy of these multi-layers in an actual detector, we fabricated a photon counting MKID from a tri-layer film with $T_C = 1.3$~K. The device design is identical to our previous experiment on a detector made from a sub-stoichiometric film \cite{Gao_SubStoichometricTiN_arXiv}. Figure 3 is a histogram of the optimally filtered pulse height (normalized by the template pulse) using the frequency readout is shown in blue. It is generated from a total of 10,000 laser pulsing events (pulse width of 200 ns and repetition rate of 1~kHz). The first 2 peaks are clearly resolved, and correspond to the events of 0 and 1 photons being absorbed in the detector. A 3-peak Gaussian profile is used to fit the data and shown by the red curve. From the fitting result we have extracted the energy resolving power $\Delta E/E = 0.6$ and the energy resolution $\Delta E = 0.48$~eV. The recombination time was 50$\mu$s. The experimental setup and measurement procedures are identical to our previous photon-counting experiment with sub-stoichiometric TiN LEKID and the resonance frequencies are within 1\%, illustrating the close similarities in kinetic inductance and other electrical properties of the two materials. This is further supported by the fact that we have observed that the tri-layer MKID used in this study also shows the anomalous temperature and pulse response as reported earlier for detectors made from substoichiometric TiN films.

In conclusion, we have utilized TiN/Ti/TiN multi-layers to create highly uniform films on 75 mm Si wafers with fine control of the $T_C$ for use as MKIDs and photon detectors. Compared to sub-stoichiometric films, the uniformity in $T_C$ is substantially improved, with much greater reproducibility between wafers. The desirable properties of TiN, i.e. high resistivity and kinetic inductance, low RF loss and superior photon detection sensitivity are found to be preserved. The tri-layer MKID is sensitive to single photons at 1550 nm and has achieved a energy resolution of $\Delta E =$0.48 eV.  For MKID applications that would require a film thicker than 20nm, the 10-layer superlattice increases the thickness to 60 nm. In addition, the 1.3 K TiN can be used as a direct replacement for Al films in current MKID designs for an immediate enhancement of sensitivity.  The fine control of $T_C$ and the uniform response across the wafer would also be interesting for nanowires and TES. Furthermore the ability of maintaining the low loss TLS interfaces implies that the range of high performance resonator devices could be extended astronomical, quantum information, and other applications.

Acknowledgements:
We acknowledge support for this work from DARPA, the Keck Institute for Space Studies, the NIST Quantum Initiative, and NASA under Contract No. NNH11AR83I. The authors thank  Jonas Zmuidzinas, Henry Leduc, and Martin Weides for helpful discussions and insights. This work is a contribution of U.S. Government, not subject to copyright

\begin{table}
\begin{tabular}{|l|l|l|l|l|l|}
\hline
type	&film				&$T_C$ [K] 	& $Q_i$ & $\lambda$ [nm]& $\rho_n$ [$\mu\Omega cm]$\\ \hline
CPW	&tri-layer 15/10/15		&2.5		&	&785*	&100 \\        
CPW	&tri-layer 4/10/4		&1.4		&250,000	& 	800&110 \\
LEKID	&tri-layer	4/10/4		&1.4		&100,000	&950	&100\\
CPW	&superlattice	3/10/$\cdot\cdot\cdot$/3	&1.4		&			&1010*	&120\\
LEKID	&sub-stoichiometric 	&1.3		&300,000	&1000	&100\\        
CPW	&tri-layer	2/15/2		&1.1		&80,000		&	900 &130\\
CPW	&tri-layer	1/20/1		&0.8		&		&	1110* &90\\
\hline
\end{tabular}
\caption{Low temperature properties of RF resonators from TiN/Ti multi-layers and a sub-stoichiometric film. For films that were not patterned into resonators, no $Q_i$ data are available, and penetration depths (marked with an asterisk) were calculated from the resistivity measured at 4 K.  }

\label{ComparisonTable}
\end{table}

\begin{figure}[b]
\includegraphics[width=6cm]{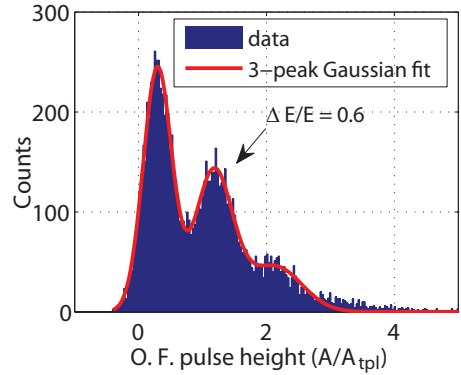}
\caption{Photon counting statistic of the 1.3~K trilayer LEKID in response to 1550~nm photons.} 
\label{Detector}
\end{figure}

\end{document}